\documentclass{phc-proc4-auth}

 \usepackage{graphicx}

\usepackage{amssymb}

\begin{document}

\begin{frontmatter}



\title{Fermionic Ising glasses with BCS pairing interaction in the presence of a
transverse field}

\author{F. M. Zimmer and S. G. Magalh\~aes}

\address{Departamento de F\'\i sica -- UFSM,  97105-900 Santa Maria, RS, Brazil}

\begin{abstract}

In the present work we have analyzed a fermionic infinite-ranged Ising spin 
glass with a local BCS coupling in the presence of transverse field. 
This model has been obtained by tracing out the conduction electrons degrees 
of freedom in a superconducting alloy. The transverse field $\Gamma$ is applied in the 
resulting effective model. The problem is formulated in the path integral 
formalism where the spins operators are represented by bilinear combination 
of Grassmann fields. The problem can be solved by combining previous 
approaches used to study a fermionic Heisenberg spin glass and a Ising spin 
glass in a transverse field. The results are show in a phase diagram $T/J$
{\it versus} $\Gamma/J~(J$ is the standard deviation of the random coupling 
$J_{ij}$) for several values of $g$ (the strength of the pairing interaction). 
For small $g$, the line transition $T_c(\Gamma)$ between the normal 
paramagnetic phase and the spin glass phase decreases when increases 
$\Gamma$, until it reaches a quantum critical point. For increasing $g$, a PAIR 
phase (where there is formation of local pairs) has been found which 
disappears when is close to $\Gamma_c$ showing that the transverse field 
tends to inhibited the PAIR phase.
\end{abstract}

\end{frontmatter}

Spin glass-like phase has been found in several physical systems such as cuprate
superconductors [1], heavy fermions [2] and conventional
superconductors [3] doped with magnetic impurities like
$Gd_xTh_{1-x}Ru_2$. Recent theoretical studies have been trying to model the
equivalent phase transition problem, at finite temperatures, between a spin
glass phase and a BCS pairing among localized fermions of opposite spins
(see Ref. [4] and references therein).
Nevertheless, disorder and frustration nearby a
quantum critical point (QCP) can be a source of non-trivial effects. 
In fact, it has been shown that there is a deviation of a Fermi-liquid behavior near $T=0$
in the transition between a metallic paramagnetic and a metallic spin glass
[5]. 

Quite recently, a quantum Ising spin glass in a transverse field
$\Gamma$ has been investigated [6]. The non-comutativity of quantum
mechanical  spins operators has been treated within the framework of path
integral formalism where both spins operators $\hat{S}_{i}^z$ and
$\hat{S}_{i}^x$ are represented by bilinear combination of Grassmann fields.
Both static approximation and the replica symmetry ``ansatz" have been used. The
results show the freezing temperature $T_f(\Gamma)$ decreases (when
$\Gamma$ increases) toward a QCP. 
 
In the present work, we studied the competition between a spin glass ordering
and BCS pair formation in a presence of a mechanism, which can lead to a
quantum phase transition. The formalism used  in the present approach is a combination of those
introduced in Ref. [4] and [6]. This has been done using a Hamiltonian with a
fermionic Ising spin glass and a BCS pairing interaction in a real space
[4] with a transverse magnetic field $\Gamma$. 
Therefore, the Hamiltonian is given by:
\begin{eqnarray}
{\hat{ H}}=-\sum_{ij}(J_{ij}\hat{S_{i}^{z}}\hat{S_{j}^{z}}+
\frac{g}{N}c_{i\uparrow}^{\dag} 
c_{i\downarrow}^{\dag}c_{j\downarrow}c_{j\uparrow})-
2\Gamma\sum_{j}\hat{S_{j}^{x}}.
\label{h1}
\end{eqnarray}
The spins operators $\hat{S_{i}^{z}}$ and $\hat{S_{j}^{x}}$ are represented 
as Ref. [6]. The coupling $J_{i j}$ is
infinite ranged with Gaussian distribuition, zero mean and variance 
$<J^{2}_{ij}>=J^{2}/2$. 

\begin{figure} 
\begin{center}
\includegraphics*[scale=.4,width=12.5pc,angle=270]{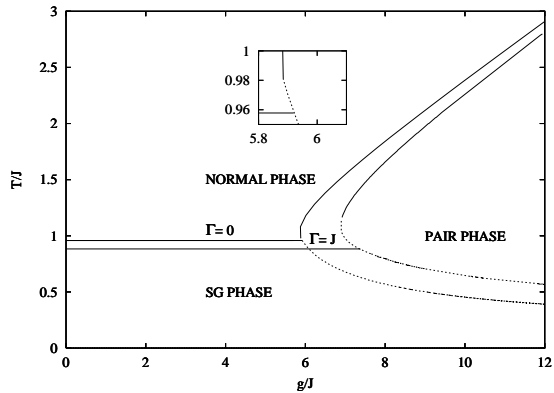}
\end{center}
\caption{Phase diagrams as a function of $T/J$ and pairing coupling $g/J$ for two
values of $\Gamma$. Solid lines indicate second-order transition while dotted
line indicate a first-order transition. The tricritical point for $\Gamma=0$ is
shown in detail in the diagram.}  
\label{fig1}
\end{figure}

The Grand Canonical potential can be obtained using the replica  trick
[4]. The action in the replicated partition function has three
components $A=A_{BCS}+A_{SG}+A_{\Gamma}$. In the first one, the $|\eta_\alpha|$ order
parameter can be introduced by a Hubbard-Stratonovich transformation following
closed the approach of Ref. [4] which
corresponds to a long range order where there is double occupation of the sites
(PAIR phase). The spin glass order parameter $q_{\alpha\beta}$ is also introduced in a similar
way, but 
in the quantum case the diagonal component of the spin glass order parameter
is no longer constrained to unity. Therefore, the resulting $A_{BCS}$ can be
written in terms of Nambu matrices (see Ref. [4]) and the
resulting spin part of the action $A_{SG}+A_{\Gamma}$  in terms of Spinors
matrices (see Ref. [6]). 
In order to solve the functional integral over the
Grassmann fields, 
the elements of Spinors and Nambu matrices have been mixed (see
Ref. [4], Eq. (20)-(22)). Further details will be given elsewhere [7].   
Therefore, the Grand Canonical potential can be found for the half-filling case
as  
\begin{eqnarray}
\frac{\Omega}{\beta}=A-\int_{-\infty}^{\infty}Dz\ln[ 
\cosh{\beta g|\eta|}+\int_{-\infty}^{\infty}Du\cosh{\beta|\bf{h}|}],
\label{eq3}
\end{eqnarray}
where $A=(\beta J)^2\bar{\chi}(\bar{\chi}+2q)/2+\beta g|\eta|^2$, ${\bf h}=J\sqrt{\theta^2+(\Gamma/J)^2}$ and $\theta=\sqrt{2\bar{\chi}}u + 
\sqrt{2q}z$. In both equation $Dz=dz\exp(-z^2/2)/\sqrt{2 \pi}$ and $\beta=1/T$
($T$ is the temperature). In $A$, $q$ is the non-diagonal 
spin glass order parameter, $\bar{\chi}=\chi/\beta$, where $\chi$ is the local suscetibility related with the 
non-diagonal spin glass order parameter.

Phases diagrams can be obtained in $T/J-g/J$ ($g$ is the pairing strength) and
$T/J-\Gamma/J$ spaces which are shown in Fig. 1 
and Fig. 2, respectively. In Fig. 1, three phases can be identified for $\Gamma=0$: a normal-paramagnetic phase (NORMAL) at high temperature and small $g$,
a spin glass phase (SG) at low temperature and enhancing $g$ one gets a phase transition at $g=g_{c}(T)$ to a PAIR phase. 
When $\Gamma$ is non-null, it can be seen that the line transition $g_{c}(T)$ is displaced showing that the existence of the PAIR phase requires 
greater values of 
$g$, at the same time the freezing temperature $T_f$ decreases towards zero. These changes with increasing $\Gamma$ can be better seen in
Fig. 2,  for $g=0$ there is just a transition line between the NORMAL and the SG phases with a QCP at $\Gamma_{c}=2\sqrt{2}J$ as already found
in Ref. [6]. If $g$ is
turned on, which energetically favors the double occupation, the PAIR phase disappears when $\Gamma$ is increased. 
The sequence Fig 2.b-Fig 2.d shows that it is necessary even greater values of $g$ in order to PAIR phase starts to occupy
a larger region than the SG phase. The QCP remains the same as $g=0$ and the replica symmetric solution is unstable in the whole SG phase.  

To conclude, our results suggest that the competition between the SG and the PAIR phases is 
strongly affected by the spin flipping mechanism ($\Gamma$) which tends to inhibited the pair formation in the 
sites, particularly near to the QCP. 

\begin{figure} 
\includegraphics[scale=.4,width=12.5pc,angle=270]{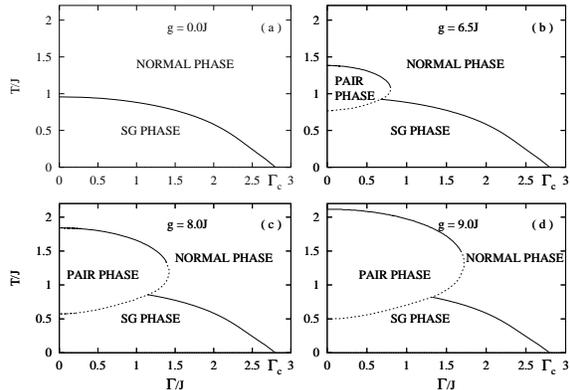}
\caption{Phase diagrams as a function of $T/J$ and $\Gamma/J$ for several fixed
values of $g/J$: (a) $g=0$, (b) $g=6.5J$, (c) $g=8J$, and (d) $g=9J$. It is used
the same convention as Fig. 1 for the transition lines.}  
\label{fig2}
\end{figure}


\end{document}